\documentclass[12pt]{article}

\usepackage{epsfig,amsmath,amssymb,graphics,graphicx} 

\newcommand{\be}{\begin{equation}}
\newcommand{\ee}{\end{equation}}
\newcommand{\bi}[1]{\vspace{-3mm} \bibitem{#1}}

\begin{document}

\today

\begin{center}
{\Large \bf 
Transition to Chaos in Discrete Nonlinear Schr{\"o}dinger Equation with Long-Range Interaction}
\vskip 5 mm

{\large \bf Nickolay Korabel$^{a,}$\footnote{Corresponding author. 
Tel.: +1 212 998 3260; fax: +1 212 995 4640. \\ E-mail address: korabel@cims.nyu.edu.}, George M. Zaslavsky$^{a,b}$} \\

\vskip 3mm

{\it $^a$ Courant Institute of Mathematical Sciences, New York University \\
251 Mercer Street, New York, NY 10012, USA, }\\
{\it $^b$ Department of Physics, New York University, \\
2-4 Washington Place, New York, NY 10003, USA }

\end{center}

\vskip 11 mm

\begin{abstract}
Discrete nonlinear Schr{\"o}dinger equation (DNLS) describes a chain of oscillators 
with nearest neighbor interactions and a specific nonlinear term. We consider its modification 
with long-range interaction through a potential proportional to $1/l^{1+\alpha}$ with fractional 
$\alpha < 2$ and $l$ as a distance between oscillators. This model is called $\alpha$DNLS. 
It exhibits competition between the nonlinearity and a level of correlation between interacting far-distanced 
oscillators, that is defined by the value of $\alpha$. We consider transition to chaos in this system as 
a function of $\alpha$ and nonlinearity. It is shown that 
decreasing of $\alpha$ with respect to nonlinearity stabilize the system. 
Connection of the model to the fractional genezalization of the NLS (called FNLS) in the long-wave approximation 
is also discussed and some of the results obtained for $\alpha$DNLS can be correspondingly 
extended to the FNLS.
\end{abstract}

\vskip 3 mm
{\small 



\vskip 3 mm


\vskip 11 mm

\section{Introduction} 

Nonlinear Schr{\"o}dinger equation (NLS) is a paradigmatic equation that describes a slowly varying 
enveloping process in the nonlinear dispersive media. Applications of the NLS has been found in almost 
all important areas of physics: nonlinear optics, plasma physics, hydrodynamics, condenced matter 
physics, biology and others. The literature on NLS is extensive and the reviews 
\cite{Kiv00,Fla98,Hen99,Kev01,Eil06,Bra98} can provide a strong impression on the importance of the subject. 
The NLS {\it per se} is an integrable system while its different types of perturbations, more related 
to the practical needs, include a broad spectrum of solutions from solitons and breathers to chaos 
and spatio-temporal turbulence. 

Between different types of perturbations one can single out two the most interesting classes: 
external time-space dependent perturbation, and discretization of the NLS presented by a kind 
of space-difference equation, called discrete NLS (DNLS). It was shown in a set of publications 
(see \cite{Cai99,Cai00,Cai01} and references therein) that specific hyperbolic structure of the 
NLS phase space leads to the transition from a soliton type dynamics to different kind of chaos, 
including spatio-temporal turbulence, under even small perturbation. 
The fact that discretization induces chaos is fairly well known and, particularly, for the DNLS it was 
studied in \cite{Herb89,Abl90,Abl93,Abl96,Abl01} and \cite{McL92,Cal96}. Destruction of solitons, 
breathers, and wave trains is similar to what occurs from external perturbations. 
A general physical mechanism of the onset of chaos induced by a discretization is known: transition 
from a differential equation to the difference one is equivalent to the appearance of a high 
frequency (in time or in space) periodic perturbation \cite{Zaslavsky}.

One can also consider DNLS as a separate problem that describes a chain of coupled 
many-particles (oscillators) with local or nonlocal interaction. 
Such a system presents a specific interest for studying transition to chaos, 
turbulence, and statistical equilibrium in many-body problem. 
A new attraction at that point is the long-range interaction (LRI) 
that was introduced in an exponential Kac-Baker form in \cite{Bak61,Kac63} 
and later in a power Lennard-Jones form in \cite{Ish82}. In another version 
of the latter case the interaction between oscillators located at the positions 
$(n,m)$, $n \ne m$, is proportional to $1/|n-m|^{1+\alpha}$. The corresponding 
DNLS will be called $\alpha$DNLS. Studying of such systems has multiple 
interest: transition to chaos and turbulence in the presence of LRI 
\cite{Maj97,Cai01a}, sinchronization in systems with 
many particles \cite{Tar05}, controling chaos \cite{Sep93}, and different manipulations 
with physical objects in optics \cite{Kiv00,Fla98a} and condenced matter \cite{Pok83,Alf}. All these 
physical features, being described by $\alpha$DNLS, are functions of $\alpha$.

For the number of particles $N = 4$ the DNLS equation can be solved exactly. 
For $N>4$ the DNLS is not integrable and chaotic solutions are possible. 
It was shown in \cite{Abl96,McL92,McL94,Herb93,Herb94,Abl97} that there are 
two main mechanisms responsible for chaotization of the DNLS. 
One of the mechanisms prevails for symmetric initial conditions 
and another for asymmetric ones. For symmetric initial conditions 
chaos was shown to emerge from the proximity to homoclinic orbits. 
For asymmetric initial conditions this mechanims do not play 
significant role, instead, perturbations induced by discretization 
and round-off errors cause random flipping of wave's direction of motion.

The primary goal of this paper is to study the transition 
to chaos in $\alpha$DNLS depending on a parameter 
$0 < \alpha < 2$ that is responsible for the appearance of 
a power-like tail in solution, i.e. for the delocalization of modes. 
Parameter $\alpha$ has a simple physical meaning: it describes a level 
of collective coupling of particles. Particularly, for $\alpha=\infty$ 
we have only nearest neighbor interaction and for $\alpha=-1$ we have 
mean field type model.

Recenly, different properties of the nonlocal DNLS were studied 
in \cite{Gai97,Joh98,Ras98,Cri01}. It was found that some properties 
of solutions depend on the interaction exponent $\alpha$ in the LRI 
case or on the exponent $\beta$ for the Kac-Baker interactions. 
Namely, for the power-law interaction it was shown that for $\alpha$ 
less than some critical value $\alpha_{cr}$, there is an interval of 
bistability where three posible stationary states exist at each value of some 
excitation number $M$ that characterize a level of nonlinearity. 
The first type of solution is a continuum-like mode, the second 
is an intrinsically localized breathing state and the third one 
is stationary itermediate state. The long-distance behavior of the intrinsically localized 
states depends on $\alpha$. For $\alpha>2$ their tails are exponential, 
while for $0<\alpha<2$ the tails are algebraic. By changing an interaction 
constant $J$, a stable solution may become unstable. A small symmetric 
force applied to the system can also trigger transigions from one type 
of stable solutions to another \cite{Joh98}. The continuum limit of this 
model was shown to be a nonlocal NLS equation \cite{Gai97}. 
Another version of the nonlocal (integro-differential) NLS equation was 
proposed in \cite{Gai96}. Unlike the usual NLS equation, this nonlocal 
NLS equation has stationary solutions only in a finite interval of 
excitation numbers $[0,M_{max}]$.

In a similar way, as dynamics of a chain of coupled oscillators can be 
reduced to the wave equation in the long wave-length limit 
$k \rightarrow 0$, the chain of $\alpha$DNLS can be reduced 
to the fractional generalization of NLS equation or Ginzburg-Landau 
equation (FGL) \cite{Wei03,Alf05,Las05,Tar05,Tar05a,Tar06}. 
It was shown in \cite{Las05,Tar05} that mapping of the 
$\alpha$DNLS equation to fractional NLS (FNLS), or similarly to 
FGL, can be realized by some transform operator. In all fractional 
equations of this type second coordinate derivative is replaced 
by the fractional Riesz derivative \cite{SKM} of order $\alpha$. 
The corresponding comparison of solutions of discrete chain of 
oscillators with the {\it sine}-potential and fractional generalization 
of the {\it sine}-Gordon equation was perfomed in \cite{Alf,Kor06}. 
The latter results of \cite{Kor06} confirm a path to the dual features of the 
systems with LRI of order $\alpha$ and the systems described 
by a corresponding equation with fractional derivative \cite{Las05,Tar05}. 

In this paper we provide a detailed study of the transition 
to chaos in $\alpha$DNLS depending on the parameter $0 < \alpha < 2$. 
It is well known that fractional values of $\alpha$ appears in 
numerous complex systems such as spin-interacting systems \cite{Lim01}, 
adatoms \cite{Pok83}, colloids \cite{Sch84}, chemical surfaces \cite{Pfe83}, 
quantum field theory \cite{Gol03,Gol04,Gol06}, etc. In correspondence 
to the results \cite{Las05,Tar05,Tar05a}, the obtained properties 
of the $\alpha$DNLS can be extended to the FNLS. It will be shown that 
onset of chaos follows as a result of competition between the nonlinearity level 
and the level of coherency of the chain of oscillators defined by the value of $\alpha$.

\section{Basic equations}

The continuous Nonlinear Schr{\"o}dinger (NLS) equation defined 
on the finite interval $[-L/2,L/2]$ with the periodic boundary 
conditions $\psi(x+L,t)=\psi(x,t)$ can be written as 
\be
\label{1}
i \frac{d\psi}{dt} + \gamma |\psi|^2 \psi + \frac{\partial^2 \psi}{\partial x^2} = 0,
\ee
where $\gamma$ is a constant ($\gamma=1$ corresponds to the focusing nonlinearity).  
In turn, the discrete nonlinear Schr{\"o}dinger equation (DNLS) 
which describes a lattice of $N$ anharmonic oscillators with 
nearest-neighbors interaction is defined as
\be
\label{2}
i \frac{d\psi_n}{dt} + \gamma |\psi_n|^2 \psi_n + \epsilon 
\left(\psi_{n+1} - 2 \psi_n + \psi_{n-1} \right) = 0, \; \; (n=1, ..., N),
\ee
where $\psi_{n+N} = \psi_n$, $\forall n$ is the periodic boundary condition. The quantity
$\psi_n = \psi_n (t)$ is the complex amplitude of the oscillator 
at site $n$. With $\epsilon = 1/(\Delta x)^2$, Eq.\ (\ref{2}) 
is seen as a standard finite difference approximation to Eq.\ (\ref{1}). 
Here $\Delta x=L/N$ is a distance between oscillators.

The Hamiltonian, $H$, and the excitation number (or norm), $M$,
\be
\label{4a}
H = \sum_{n=1}^N \left( \epsilon |\psi_{n+1} - \psi_{n}|^2  - \frac{\gamma}{2} |\psi_n|^4 \right), \; \; \;
M = \sum_{n=1}^N |\psi_n|^2,
\ee
are the conserved quantities. 

The model which we study in the following is described by the Hamiltonian
\be
\label{7}
H = \frac{1}{2} \sum_{n=1}^{N} \left(  \sum_{\substack{m=1 \\ m \ne n}}^{N}  
J_{n-m}|\psi_m - \psi_n|^2 - \gamma |\psi_n|^4 \right),
\ee
with the periodic boundary conditions: $\psi_{n+N}=\psi_n$. 
The coupling function is defined by
\be
J_{n-m}=\frac{J}{|n-m|^{1+\alpha}}, \; \; (n \ne m),
\ee  
where $J$ is a coupling constant and $\alpha$ is an exponent which, 
depending on the physical situation, can be integer or fractional. 
The standard nearest-neighbor DNLS equation is recovered in the 
limit $\alpha \rightarrow \infty$. From the Hamiltonian 
Eq. (\ref{7}) one obtain the equation of motion
\be
i \frac{d\psi_n}{dt} = \partial H/\partial \psi_{n}^{*},
\ee
\be
\label{8}
i \frac{d\psi_n}{dt} + \gamma |\psi_n|^2 \psi_n + \sum_{\substack{m=1 \\ m \ne n}}^{N} J_{n-m}(\psi_{n} - \psi_{m}) = 0, \; \; (n=1, ..., N).
\ee 

Consider the infinite chain of equidistant oscillators ($N \rightarrow \infty$ in Eq.\ (\ref{8})). 
The Fourier transform of $\psi_n$ is given as
\be
\label{f1}
\hat{\psi}_n (k,t) = 
 \sum_{n=-\infty}^{\infty} \psi_n (t) \; \exp(- i k x_n) 
\equiv \mathcal{F}_{\Delta} \{\psi_n (t)\},
\ee
where $x_n = n \Delta x$ is a coordinate of the $n$-th oscillator, 
and $\Delta x = 2 \pi/K$ is a distance between oscillators. 
Here we have treated $k$ as a continuous variable.
The inverse transform is defined as
\be
\label{f2}
\psi_n (t) = \frac{1}{K} \int_{-K/2}^{+K/2} dk \; \hat{\psi}_n (k,t) \; \exp(i k x_n) 
\equiv \mathcal{F}_{\Delta}^{-1} \{\hat{\psi}_n (t)\}, 
\ee
Transition to the continuous limit 
$\Delta x \rightarrow 0$ ($K \rightarrow \infty$) can be obtained by 
transforming $\psi_n (t) = \psi(n \Delta x,t) = \psi(x_n,t) \rightarrow \psi(x,t)$. 
Then, changing sums into integrals, Eqs. (\ref{f1}), (\ref{f2}) become
\be \label{ukt2} 
\tilde{\psi}(k,t)=\int^{+\infty}_{-\infty} dx \ e^{-ikx} \psi(x,t) 
\equiv {\cal F} \{ \psi(x,t) \}, 
\ee
\be \label{uxt}
\psi(x,t)=\frac{1}{2\pi} \int^{+\infty}_{-\infty} dk \ e^{ikx} \tilde{\psi}(k,t) 
\equiv {\cal F}^{-1} \{ \tilde{\psi}(k,t) \}, 
\ee
where
\be 
\tilde{\psi}(k,t)= {\cal L} \hat{\psi}(k,t), \quad \psi(x,t)= {\cal L} \psi_n(t) = {\cal L} \psi(x_n,t), 
\ee
and ${\cal L}$ denotes the limit $\Delta x \rightarrow 0$. 
Operation $\bar{T}={\cal F}^{-1} {\cal L} \ {\cal F}_{\Delta}$ 
can be called a transform operator (or transform map), 
since it performs a tranfrormation of a discrete model of 
coupled oscillators to the continuous media. For $0 < \alpha < 2$ 
application of $\bar{T}$ leads to the fractional NLS. 
For more details, definitions and 
applications of the transform operator to different systems 
see \cite{Tar05,Tar05a,Kor06}. In a brief form this approximation is as 
follows \cite{Las05,Tar05}. 

First we perform the Fourier tranformation of Eq.\ (\ref{8}) 
for infinite chain of oscillators ($N \rightarrow \infty$) 
\be \label{C3}
i \frac{\partial \hat{\psi}(k,t)}{\partial t} + \gamma \; 
\mathcal{F}_{\Delta} \{ |\psi_n|^2 \psi_n \} + J \; ( \hat{J}_{\alpha}(k) - \hat{J}_{\alpha}(0) ) 
\; \hat{\psi}(k,t)  = 0,
\ee
where 
\be \label{C5}
\hat{J}_{\alpha}(k) = \sum_{\substack{n=-\infty \\ n \ne 0}}^{+\infty} 
 \frac{e^{-i k n \Delta x}}{|n|^{1+\alpha}}= 
\sum^{+\infty}_{n=1} \frac{e^{-i k n \Delta x} + e^{i k n\Delta x}}{n^{1+\alpha}}   = 
Li_{1+\alpha}( e^{i k \Delta x} ) + Li_{1+\alpha}( e^{-i k \Delta x} ),
\ee
and $Li_{1+\alpha}(z)$ is a polylogarithm function \cite{Erd} with series representation 
\be \label{D2}
\hat{J}_{\alpha}(k)= a_{\alpha} \; |\Delta x|^{\alpha} \; |k|^{\alpha} +
2 \sum^{\infty}_{n=0} \frac{\zeta(1+\alpha-2n)}{(2n)!} (\Delta x)^{2n} (-k^2)^n , 
\quad |k|< 1, \quad \alpha \not=0,1,2,3 ... \; .
\ee
Here $J_{\alpha}(0)=2 \zeta(1+\alpha)$, $\zeta$ is the Riemann zeta-function and
\be \label{D5}
a_{\alpha} =  2 \; \Gamma(-\alpha) \; \cos \left( \frac{\pi \alpha}{2} \right).
\ee 
For $\alpha=2$, $\hat{J}_{\alpha}(k)$ reduces to the 
Clausen function $Cl_{2}(k)$ \cite{Lew81}.

Substitution of Eqs. ({\ref{D2}}) and ({\ref{D5}}) into (\ref{C3}) gives
\be \label{D4}
i \frac{\partial \hat{\psi}(k,t)}{\partial t}  + 
J \; \hat{\psi}(k,t) \left( \; a_{\alpha} |\Delta x|^{\alpha} \; 
|k|^{\alpha} + 2 \sum^{\infty}_{n=0} \frac{\zeta(\alpha+1-2n)}{(2n)!} 
(\Delta x)^{2n} (-k^2)^n - \hat{J}_{\alpha}(0) \right) +
\ee
\[ + \gamma \; \mathcal{F}_{\Delta} \{ |\psi_n|^2 \psi_n \} = 0.\]
Note that $\hat{J}_{\alpha}(0)$ exactly cancels the constant which 
is the first term of the sum in Eq.\ (\ref{D4}). 
In the limit $k \rightarrow 0$ Eq.\ (\ref{D4}) yields
\be \label{Appr}
i \frac{\partial \hat{\psi}(k,t)}{\partial t} + \bar{J} \; 
\hat{\mathcal{T}}_{\alpha, \Delta}(k) \; \hat{\psi}(k,t) + \gamma \; 
\mathcal{F}_{\Delta} \{ |\psi_n|^2 \psi_n \} = 0, 
\ee
where $\bar{J}=J |\Delta x|^{min\{\alpha,2\}}$ and
\be \label{D8}
\hat{\mathcal{T}}_{\alpha, \Delta}(k) = \begin{cases} 
a_{\alpha} |k|^{\alpha} - |\Delta x|^{2-\alpha} \zeta (\alpha -1) k^2, & 0 < \alpha < 2, \quad (\alpha \not=1) 
\cr  
|\Delta x|^{\alpha-2} a_{\alpha} |k|^{\alpha} - \zeta (\alpha -1) k^2, & 2< \alpha < 4, \quad (\alpha \not=3).
\end{cases}
\ee
For $0 < \alpha < 2$ the operator $\hat{\mathcal{T}}_{\alpha, \Delta}(k)$ is defined up to 
$O(k^2)$ and for $2 < \alpha < 4$ up to $O(|k|^{\alpha})$. 

Eq.\ (\ref{D8}) has a crossover scale for 
\be \label{D10}
k_0 = |a_{\alpha}/\zeta (\alpha-1)|^{1/(2-\alpha)} |\Delta x|^{-1}.
\ee
From Eq.\ (\ref{D10}) it follows that 
$\hat{\mathcal{T}}_{\alpha, \Delta}(k) \sim k^2$ for $\alpha>2$, $k \ll k_0$ 
and nontrivial expression 
$\hat{\mathcal{T}}_{\alpha, \Delta} (k) \sim |k|^{\alpha}$ 
appears only for $\alpha<2$, $k \ll k_0$. This crossover was 
considered also in \cite{Las05,Tar05}. 
Performing the transition to the limit $k \ll k_0 $ 
(or more precisely $k \Delta x \ll k_0 \Delta x$) 
and applying the inverse Fourier transform to (\ref{Appr}), we obtain
\be \label{D10b}
i \frac{\partial}{\partial t} \psi(x,t) + \bar{J} \; \mathcal{T}_{\alpha}(x) \; \psi(x,t) + \gamma \; |\psi(x,t)|^2 \psi(x,t) = 0 \quad \alpha \not=0,1,2,...,
\ee
where 
\be \label{Tx} 
\mathcal{T}_{\alpha}(x) = 
\mathcal{F}^{-1} \{ \hat{\mathcal{T}}_{\alpha} (k) \} = 
\begin{cases} 
- a_{\alpha} \frac{\partial^{\alpha}}{\partial |x|^{\alpha}} , & 0 < \alpha < 2, \quad (\alpha \not=1) \cr 
\zeta (\alpha -1) \frac{\partial^2}{\partial x^2}, & \alpha > 2, \quad (\alpha \not=2,3,4,...);
\end{cases}
\ee
\[
\hat{\mathcal{T}}_{\alpha} (k) = 
\begin{cases} 
a_{\alpha} |k|^{\alpha}, & 0 < \alpha < 2, \quad (\alpha \not=1) \cr 
- \zeta (\alpha -1) \; k^2, & \alpha > 2, \quad (\alpha \not=2,3,4,...).
\end{cases}
\] 
Here, we have used the connection between the Riesz fractional derivative and its Fourier transform \cite{SKM}: 
\be
|k|^{\alpha} \longleftrightarrow - \frac{\partial^{\alpha}}{\partial |x|^{\alpha}}, \quad k^2 \longleftrightarrow - \frac{\partial^2}{\partial x^2}.
\ee
Properties of the Riesz fractioanl derivative can be found in \cite{SKM,OS,MR,Podlubny}. 

In the following section we will consider simulations of the finite 
chain of oscillators on a finite interval $(-L/2,L/2)$ ($n=1, ..., N$, where $N$ is even) 
described by the Hamiltonian
\be
\label{haml}
H = \sum_{n=1}^{N} \left( J \sum_{l=1}^{N/2-1} \frac{|\psi_{n+l} - \psi_n|^2}{l^{1+\alpha}} - 
\frac{\gamma}{2} |\psi_n|^4 \right)
\ee 
and equation of motion 
\be
\label{haml2}
i \frac{d\psi_n}{dt} + \gamma |\psi_n|^2 \psi_n + J \sum_{l=1}^{N/2-1} 
\frac{\psi_{n+l} - 2 \psi_n + \psi_{n-l}}{l^{1+\alpha}} = 0, \; \; (n=1, ..., N).
\ee
The form of the Hamiltonian and the equation of motion account for the periodic 
boundary conditions $\psi_{n+N} = \psi_n$. To avoid double counting of interactions 
we introduce a cutoff for the interaction range at $N/2-1$.

All initial conditions and 
parameters are choosen to make it possible to use the transform 
operator $\bar{T}$. This means that for not too large time 
the obtained results can be also applied to the FDNLS equation (\ref{D10b})
\be
i \frac{\partial}{\partial t} \psi(x,t) - G \; 
\frac{\partial^{\alpha} \psi(x,t)}{\partial |x|^{\alpha}} \; 
\psi(x,t) + \gamma \; |\psi(x,t)|^2 \psi(x,t) = 0, \; \quad 0 < \alpha < 2, \quad \alpha \not=1,
\ee
\be
G = 2 \; J \; |\Delta x|^{\alpha} \; \Gamma (-\alpha) \cos (\pi \alpha/2).
\ee
It is also worthwile to mention that for $\alpha \rightarrow 2$ 
we satisfy the conditions of the continuous limit approach 
since $k_0 \rightarrow \infty$, that means the corresponding 
results should be close to the solution of NLS. 
Thus, considering $\alpha$ far from $\alpha=2$, one can compare 
the solutions of $\alpha$DNLS with the solutions of 
DNLS with the nearest neighbor interaction. Such results provide a role 
of the long-range interaction for the chain dynamics. 

\section{Numerical results}

In this section numerical results obtained from solutions of the equations of motion 
Eq.\ (\ref{haml2}) on the finite interval $(-L/2,L/2)$ are summarized. 
Parameter $J$ was normalized to $J/J_0$, where 
\be 
J_0 = \sum_{n=1}^{N/2-1} \frac{1}{|n|^{1+\alpha}}. 
\ee
For all sets of parameters we integrate the equations of 
motion Eq. (\ref{haml2}) up to the time $T=100$. Numerical solutions 
were stored at each $t_q = qT/Q$, $q=0, .. ,Q-1$, where $Q=10^3$. 
The number of oscillators was $N=32$ in all simulations.

To visualise the numerical solution we plot the surface 
$|\psi(x,t)|^2$ and the "phase portrait" of the central oscillator ($n=0, x_n = 0$) 
formed by the variables:
\be
\label{Aplate}
A(t) = |\psi(0,t)|^2, \; \; A_t = dA/dt.
\ee 
Also we plot $Im(\psi(0,t))$ vs. $Re(\psi(0,t))$ and the phase 
difference of two nearby trajectories $\psi(0,t)$, $\psi(\Delta x,t)$
\be
\label{dfplate}
df = \tan^{-1}(Re(\psi(0,t))/Im(\psi(0,t))) - \tan^{-1}(Re(\psi(\Delta x,t))/Im(\psi(\Delta x,t))),
\ee
where $\Delta x=L/N$. We also calculate the discrete Fourier 
transform of the sequence $\psi(0,t_q)$, which is defined 
\be
\label{f_time}
\hat{\psi}_n(w_j)  = \frac{1}{Q} \sum_{q=0}^{Q-1} \psi_n(t_q) \; \exp(- i \; w_j \; t_q),
\ee
\be
\label{f2_time}
\psi_n (t_q) = \sum_{j=0}^{Q-1} \hat{\psi}_n (w_j) \; \exp(i \; w_j \; t_n), 
\ee
where the wavenumber is $w_j=2 \pi j/Q$, $j=0, ..., Q-1$. 
The corresponding power spectrum $S$ of the sequence $\psi_n(t_q)$ ($q=0, ..., Q-1$) is given by 
\be
\label{Splate}
S_j \equiv S(w_j) = |\hat{\psi}_n (w_j)|^2.
\ee
\begin{figure}
\centering
\rotatebox{0}{\includegraphics[width=14 cm,height=14 cm]{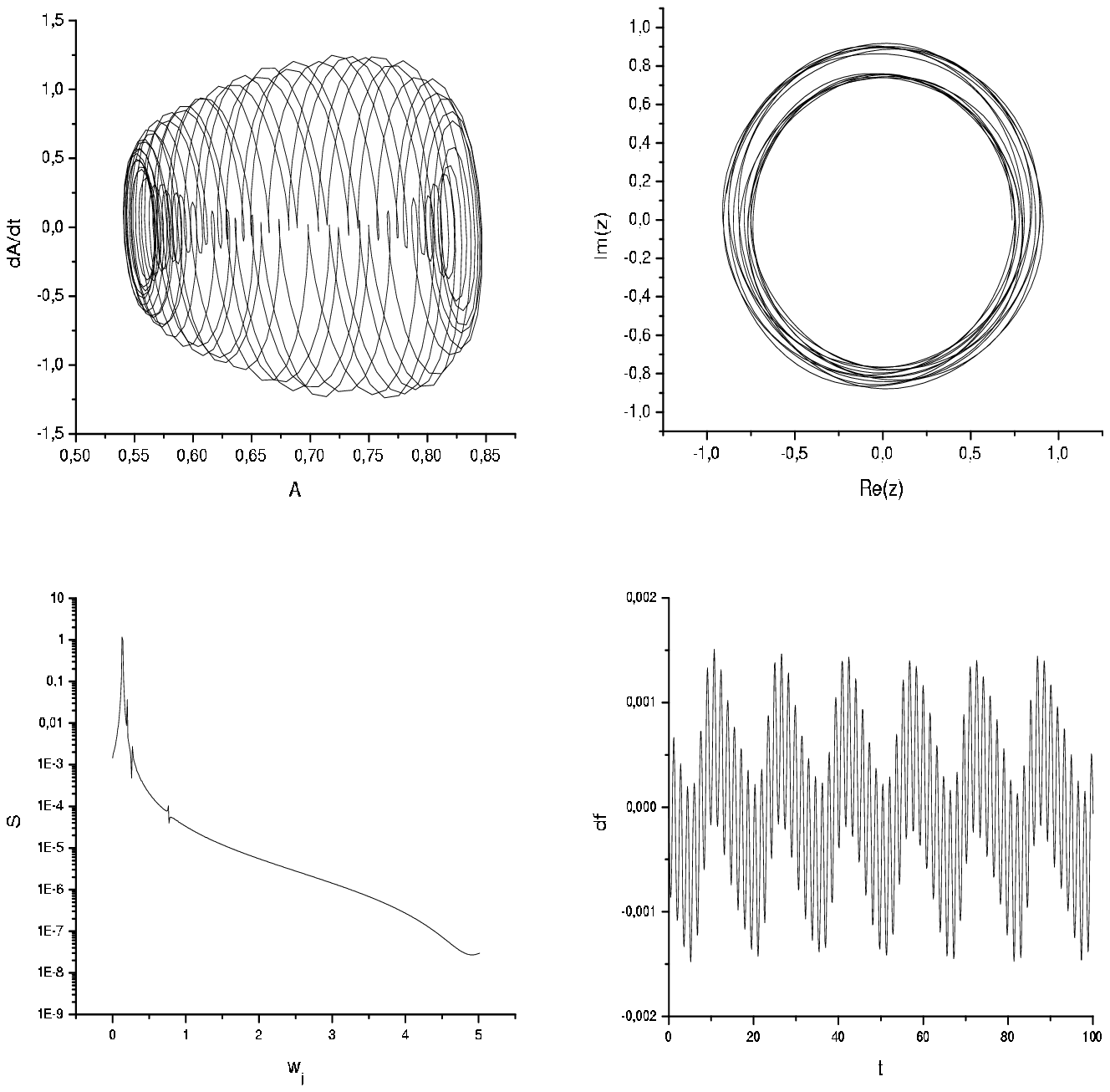}}
\caption{\label{fig1} 
Time evolution of the central oscillator. The values of parameters are 
$\alpha=1.11$, $J/J_0=0.7$ and $M=12.5$. The initial condition is given by Eq.\ (34). 
}
\end{figure}
\begin{figure}
\centering
\rotatebox{0}{\includegraphics[width=14 cm,height=14 cm]{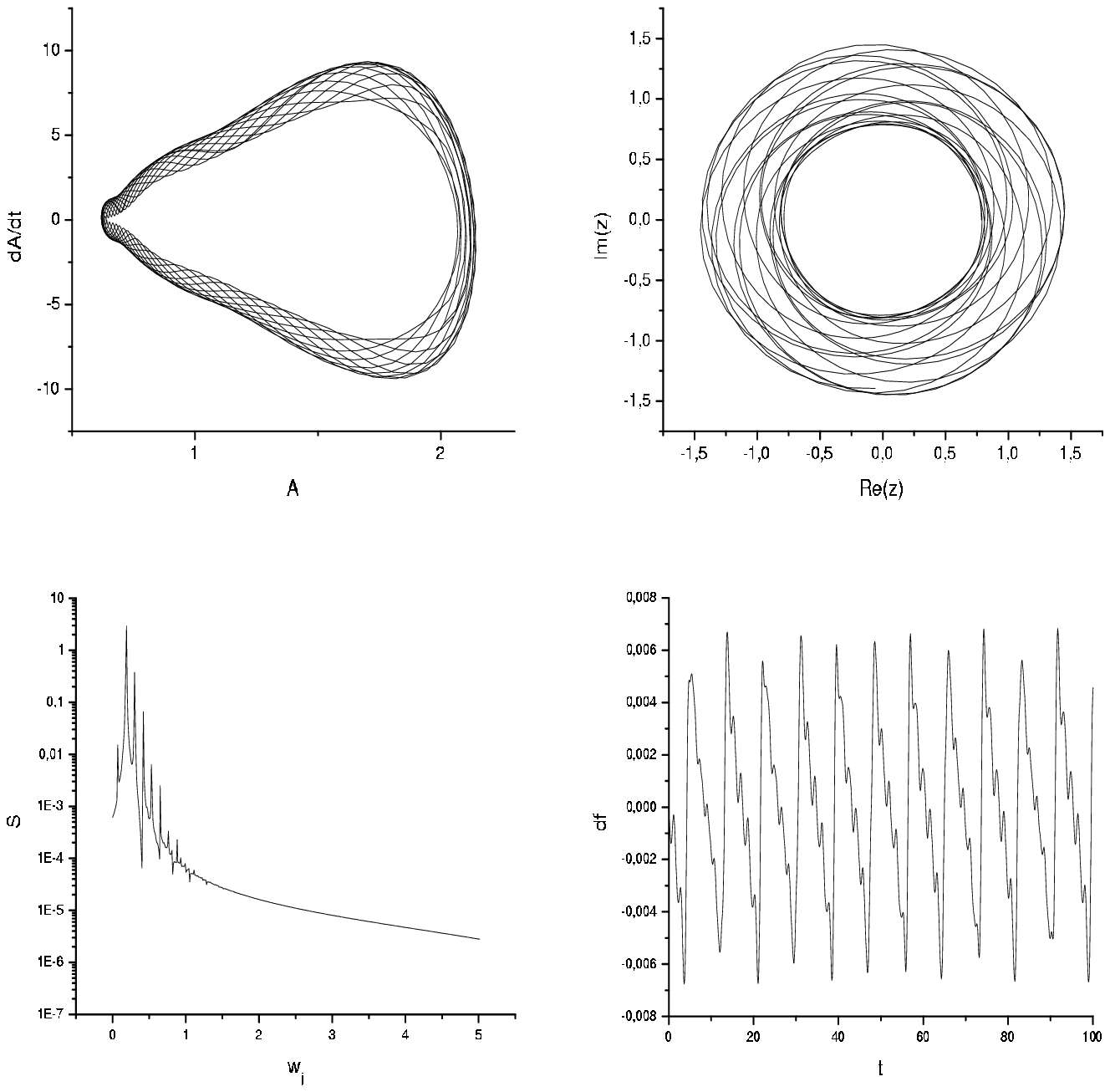}}
\caption{\label{fig2} 
Time evolution of the central oscillator. The values of parameters are 
$\alpha=1.11$, $J/J_0=0.7$ and $M=14.28$. The initial condition is given by Eq.\ (34). 
}
\end{figure}
\begin{figure}
\centering
\rotatebox{0}{\includegraphics[width=14 cm,height=14 cm]{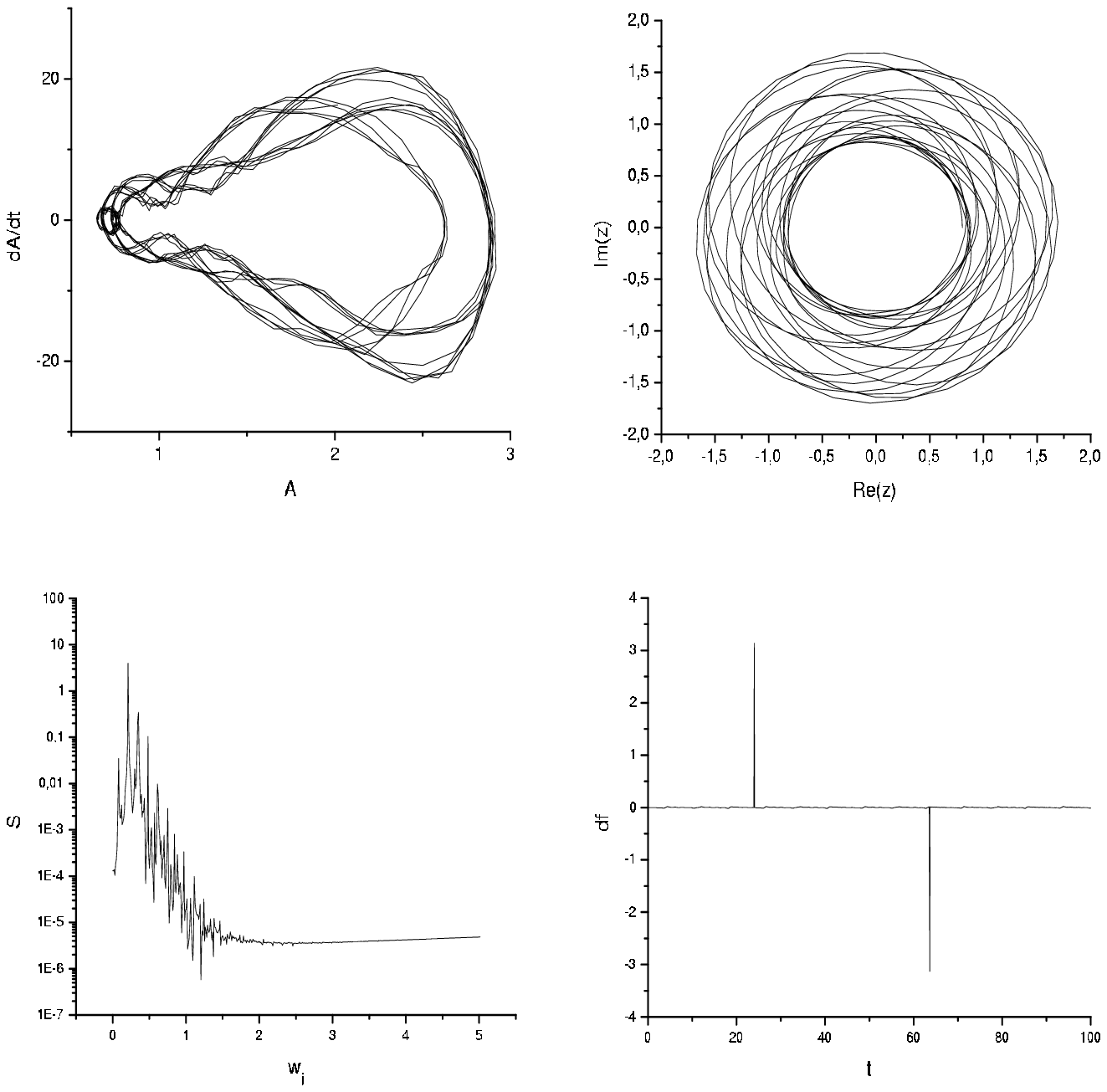}}
\caption{\label{fig3}
Time evolution of the central oscillator. The values of parameters are 
$\alpha=1.11$, $J/J_0=0.7$ and $M=14.92$. The initial condition is given by Eq.\ (34). 
}
\end{figure}
\begin{figure}
\centering
\rotatebox{0}{\includegraphics[width=14 cm,height=14 cm]{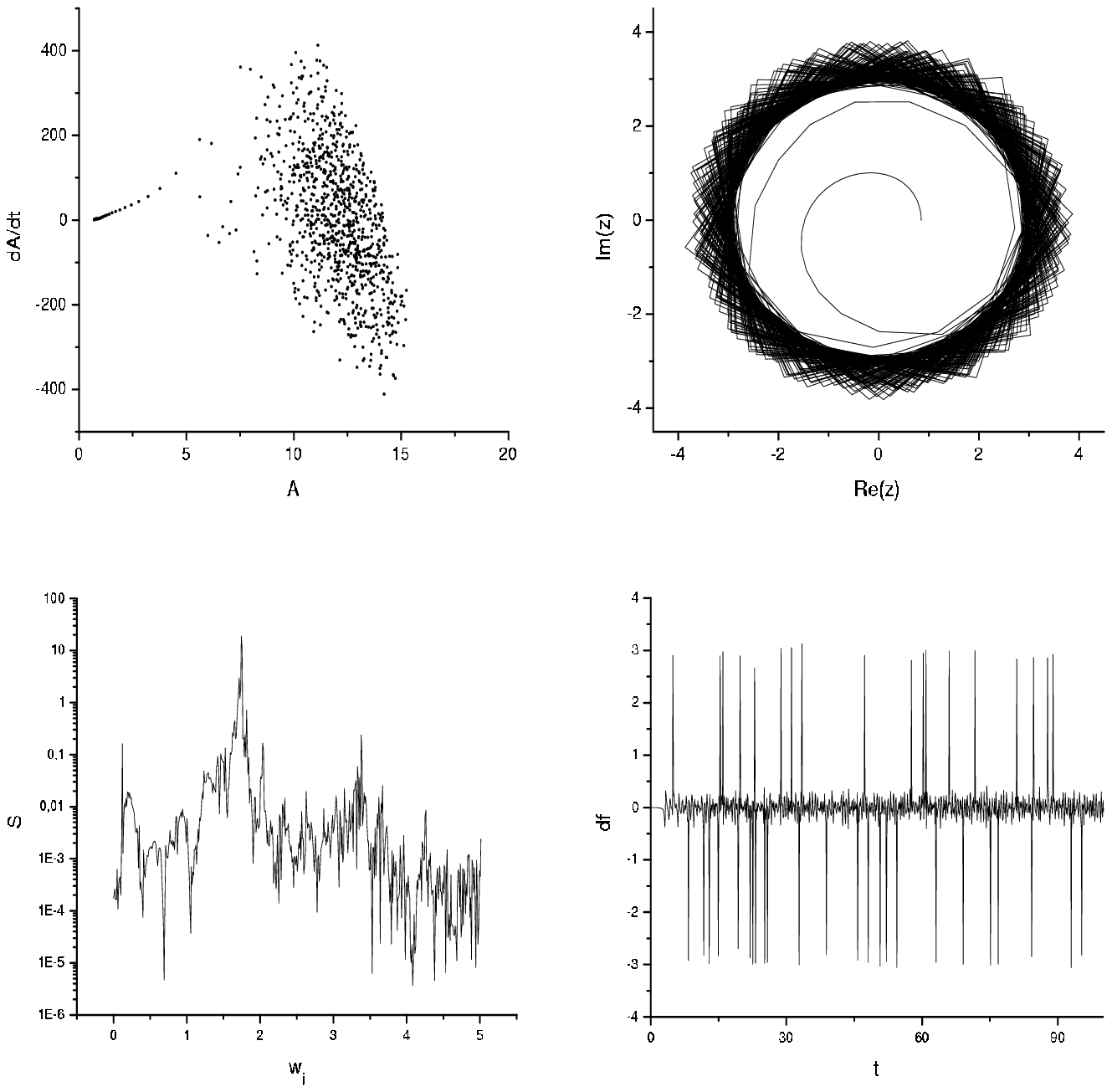}}
\caption{\label{fig4} 
Time evolution of the central oscillator. The values of parameters are 
$\alpha=1.11$, $J/J_0=0.7$ and $M=16.6$. The initial condition is given by Eq.\ (34). 
}
\end{figure}
\begin{figure}
\hspace*{-2cm}
\rotatebox{0}{\includegraphics[width=18 cm,height=24 cm]{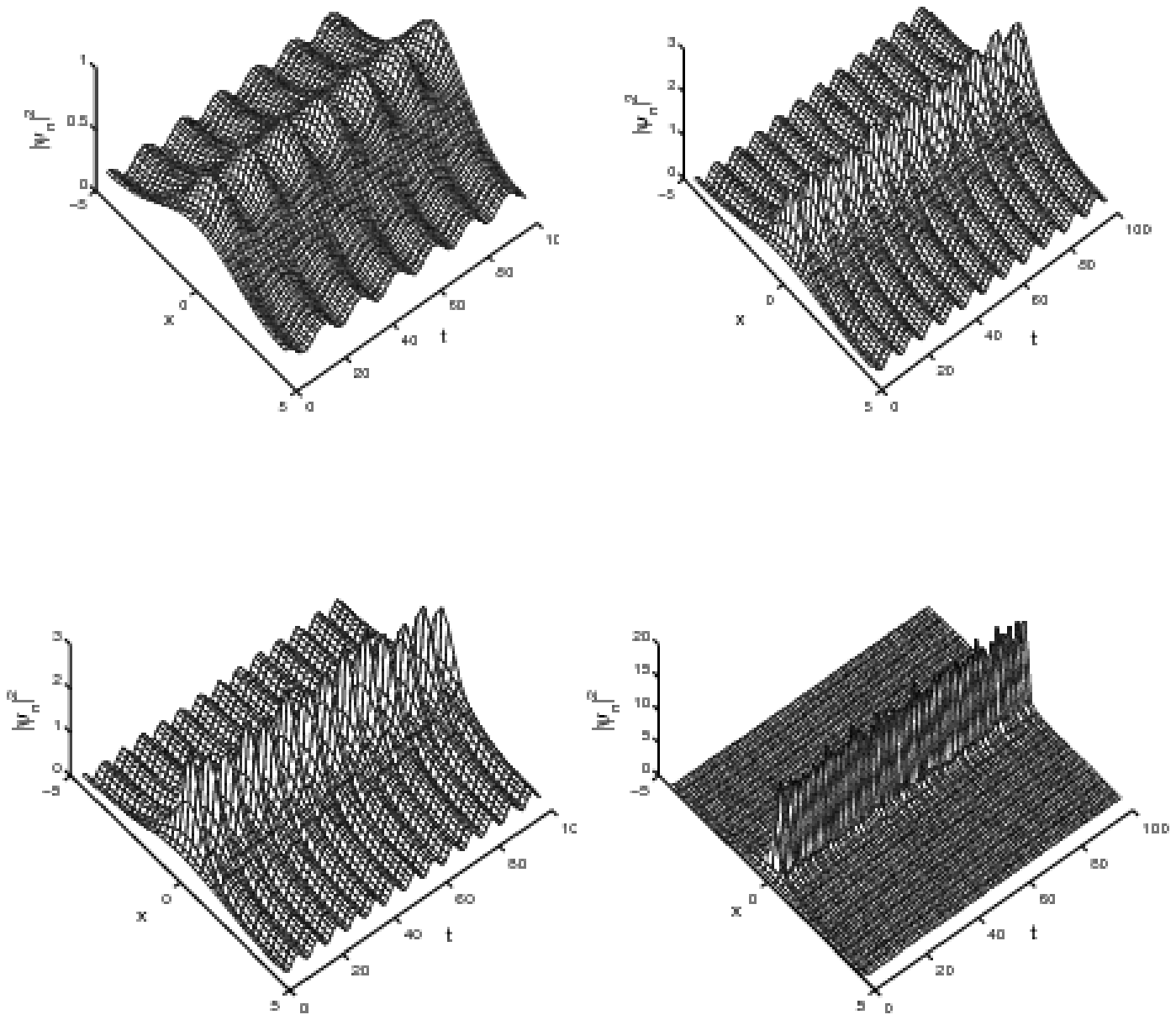}}
\vspace*{-5cm}
\caption{\label{5} 
Surfaces $|\psi_n(t)|^2$ for the same values of parameters as in Figs.\ 1-4:  
$\alpha=1.11$, $J/J_0=0.7$ and $M=12.5$ (top left), $M=14.28$ (top right), 
$M=14.92$ (bottom left) and $M=16.6$ (bottom right).}
\end{figure}

In our simulation we use two types of initial conditions: 
symmetric and asymmetric types. Symmetric initial conditions are defined as 
\be
\label{con1}
\psi(x,0) = a + b \cos \left( \frac{2 \pi}{L} x \right), 
\ee
with constants $a=0.5$, $b = 0.1$ and $L=2 \sqrt{2} \pi$. 
This form of symmetric initial conditions was also used 
in \cite{McL92, McL94} to study chaos in discretizations of standard NLS equation. 
For asymmetric initial conditions we use the following expression 
\be
\label{con2}
\psi(x,0) = a \left[ 1 + b \left\{ e^{i c} \cos 
\left(\frac{2 \pi}{L} x\right) + e^{i d} \sin \left(\frac{2 \pi}{L} x\right) \right\} \right],
\ee
where $a=1$, $b=0.2$, $c=0.9$, $d=2.03$ and $L=2 \sqrt{2} \pi$. The form of the initial condition 
is taken the same as in \cite{Abl96}. 

Our main goal is to compare solutions of Eq.\ (\ref{haml2}) for 
different values of $\alpha \in (1,2)$ and consider transition to chaotic 
dynamics of the chain of oscillators as a function of $\alpha$. 
The larger is $M$, the stronger is nonlinearity. The larger is $\alpha$, 
the smaller is LRI. The main differences in the physical properties of the 
oscillators dynamics are defined by a competition between $\alpha$ and $M$. 
In all simulations we fix $J/J_0 = 0.7$ as the value close to the one 
considered in \cite{Gai97}. All plots will show the following properties of the 
oscillators: plane ($dA/dt, A$) shows projection of the trajectory of the 
central oscillator in phase space (see definition in (\ref{Aplate})); 
plane ($Im\ z, Re\ z$) shows projection of the complex amplitude $z=\psi(0,t)$ 
of the central oscillator as a function of time; phase difference with the 
adjacent oscillator to the central one (see Eq.\ (\ref{dfplate})); spectrum of time oscillations of 
$\psi(0,t)$ (see definition in Eq.\ (\ref{Splate})); and surfaces $|\psi_n(t)|^2$ vs $t$ and $n$.

First four Figures 1-4 aim to show different regimes of the 
chain of oscillators behavior for $\alpha = 1.11$.
For small values of $M$ solutions are quasi-periodic in time 
with only few modes in the Fourier spectrum. This behaviour for 
$M=12.5$ is shown on Fig.\ 1. The plane ($dA/dt, A$), 
the plane $(Im\ z, Re\ z)$ and the phase difference 
plane demonstrate quasi-periodic behavior. 
As we increase the value of $M$, the spectrum is broadening. 
Figure 2 demonstrates the behaviour of the system with 
$\alpha=1.1$, $J/J_0=0.7$ and $M=14.28$. The quasi-periodic structure 
of the plane ($dA/dt, A$), the plane $(Im\ z, Re\ z)$ 
and the phase difference plane get more complex. 
More and more Fourier modes appear in the spectrum. This is even more 
pronounced for the case $M=14.92$ depicted in Fig.\ 3, which can be 
considered as a begining of chaos. The phase difference 
of two nearby oscillators shows two 'flips' to $\pi$ and $-\pi$ which 
indicates phase decoherence and transition to chaos. In Fig. 4 for 
$M=16.6$ the phase difference of two nearby oscillators has many 'flips' 
to $\pi$ and $-\pi$ and the Fourier spectrum of $\psi(0,t)$ becomes broad 
what is typical for chaotic dynamics. Surfaces 
$|\psi(x,t)|^2$ for the cases of Figs.\ 1-4 are 
shown in Fig.\ 5. In addition to the chaotization of the 
solution which was described above, it is seen that with increasing values 
of $M$, surfaces become more localized around $x=0$, oscillations in the 
wings of the solutions gradually decrease and the aplitude of the solution 
is increased. This can be explained by the growth of the nonlinear term with 
the growth of $M$. The role of a nonlinear coupling becomes more important 
than coherent connection of oscillators due to the LRI. 
\begin{figure}
\centering
\rotatebox{0}{\includegraphics[width=14 cm,height=14 cm]{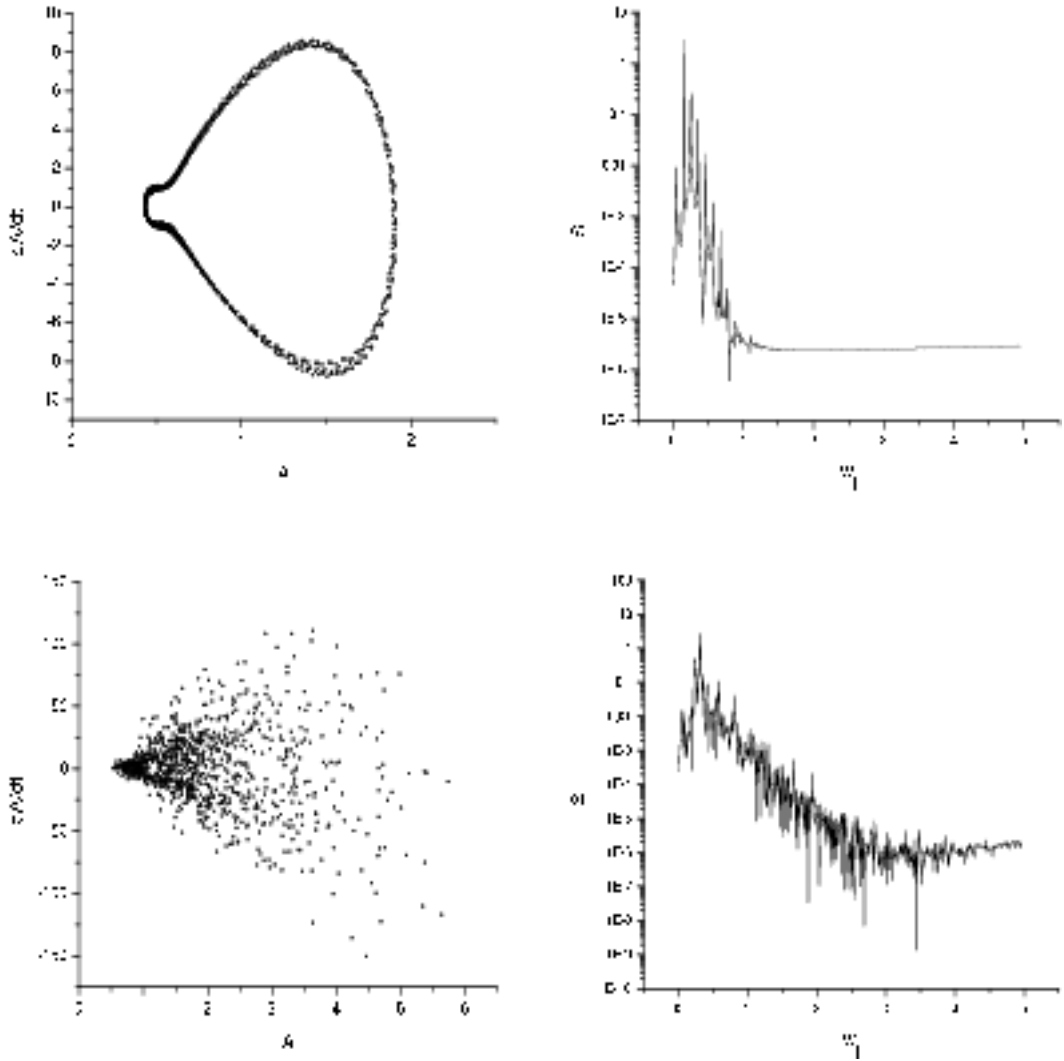}}
\caption{\label{fig6} 
The phase plane of $\psi(0,t)$ and the power spectrum $S$ of $\psi(0,t)$ for 
parameters $\alpha=1.91$, $J/J_0=0.7$, $M=10$ (top row) and $M=12.5$ (bottom row). 
}
\end{figure}
\begin{figure}
\hspace*{-2.5cm}
\rotatebox{0}{\includegraphics[width=20 cm,height=26 cm]{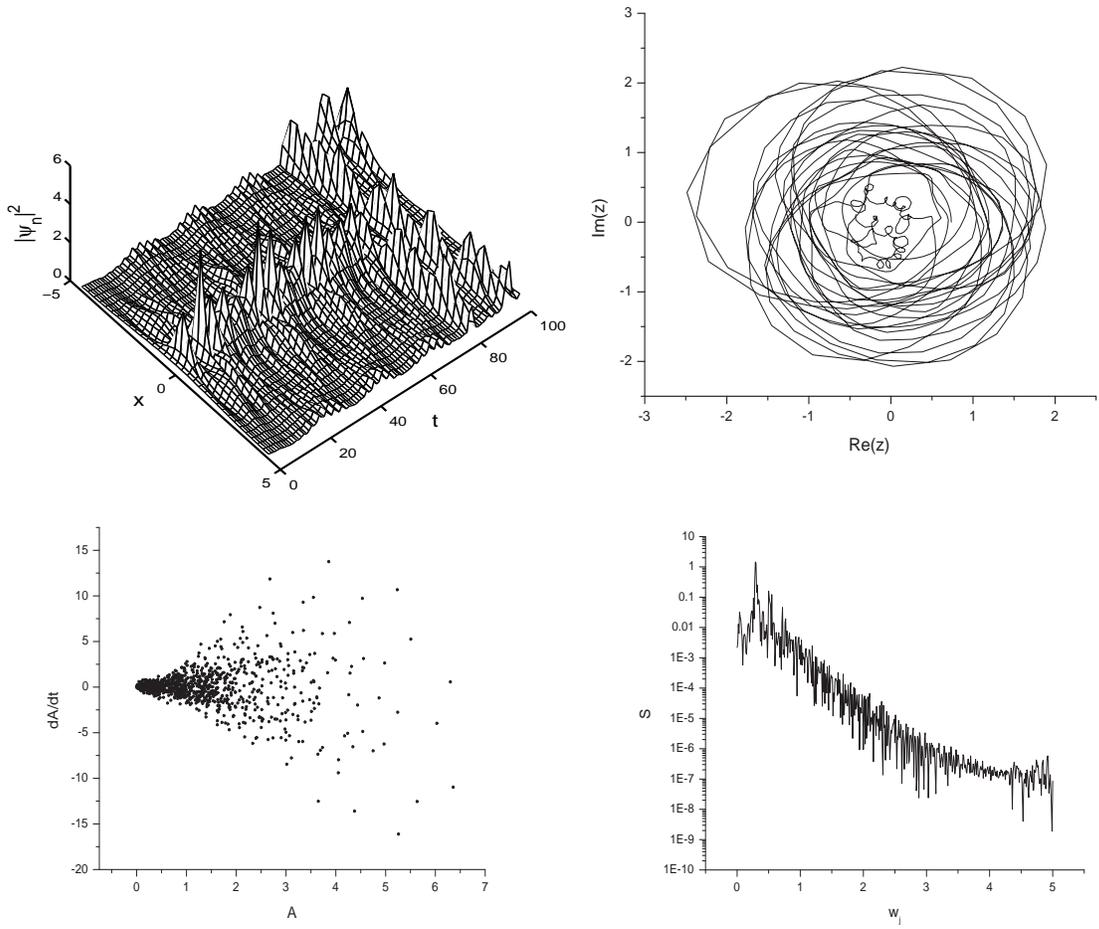}}
\vspace*{-8cm}
\caption{\label{fig7} 
Solution of the standard DNLS equation 
with $M=12.5$ and $J/J_0=0.7$. The initial condition given by Eq.\ (34).
}
\end{figure}
\begin{figure}
\hspace*{-2.5cm}
\rotatebox{0}{\includegraphics[width=20 cm,height=26 cm]{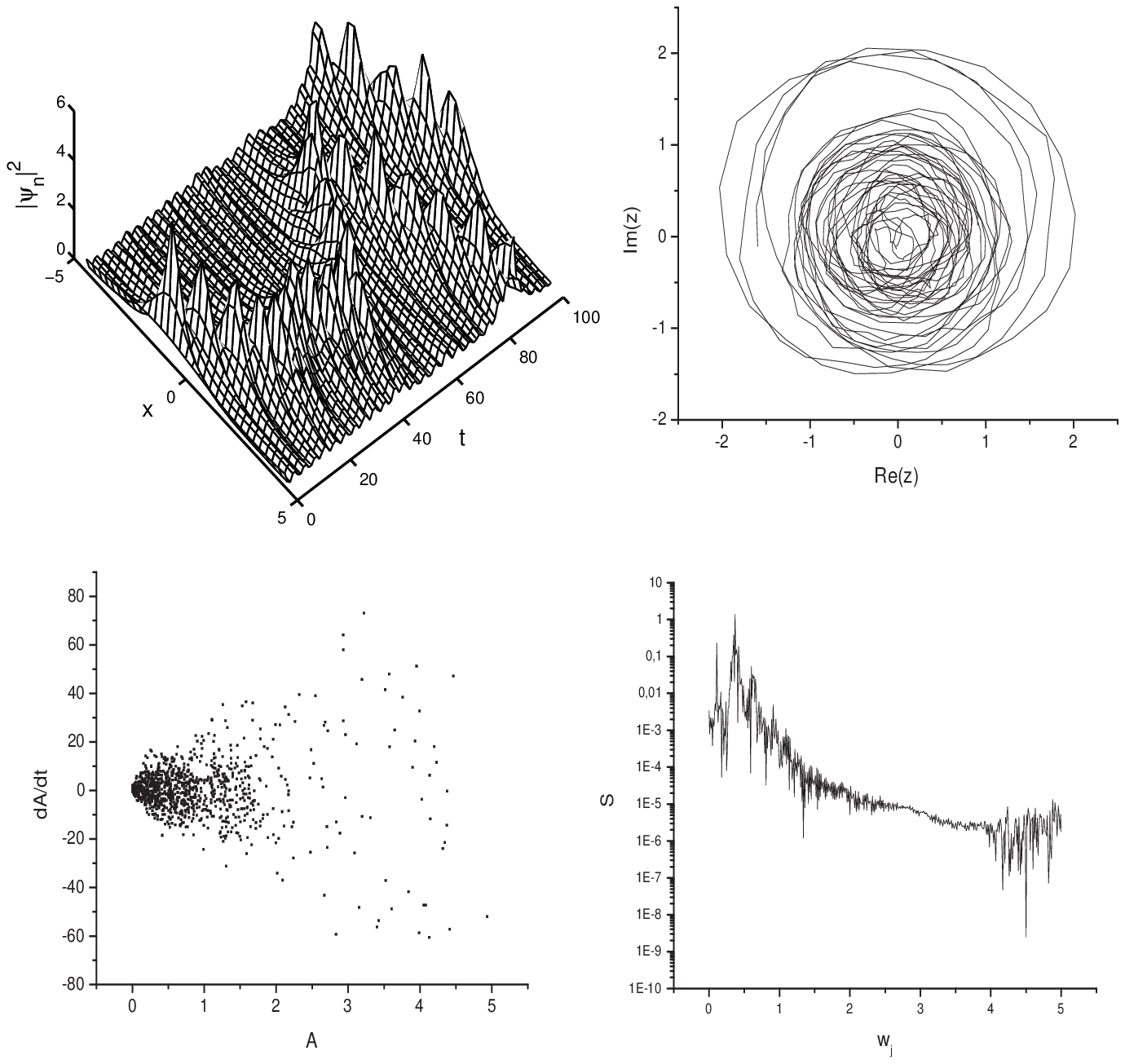}}
\vspace*{-8cm}
\caption{\label{fig8} 
Time evolution of the system of coupled oscillators with LRI and asymmetric initial conditions. 
The values of parameters are $\alpha=1.11$, $M=22.22$, $J/J_0=1$. 
The initial condition is given by Eq.\ (35).}
\end{figure}

Increasing of $\alpha$ leads to appearence 
of chaotic dynamics for smaller $M$, without significant changes 
of the diagrams and planes shown in Figs. 1-5. This is demonstrated for 
$\alpha = 1.91$ in Fig.\ 6. For $\alpha=1.51$ chaos starts at approximately $M=14.28$ and for 
$\alpha=1.91$ at $M=12.5$. Dynamics for $\alpha < 1$ is approximately similar to 
the dynamics for $1 < \alpha < 2$ and transition to chaos for $\alpha=0.73$ occurs at $M \sim 18$.
The growth of $\alpha$ leads to the increasing 
of the energy of oscillations in the tails of the solutions. 
Transition to the strongly developed chaos is not too sharp in time.

All these results could be compared to the standard DNLS equation 
with the nearest-neighbor interaction. Solution of this equation 
with the symmetric initial condition Eq.\ (\ref{con1}) is shown in 
Fig.\ 7 for $M=12.5$. There are several main distinctions of this case from the transition 
to chaos when $\alpha < 2$: (i) the symmetry of the solution breaks 
down for larger $M$ and sharply in time; (ii) the spatial chaos occurs visually 
earlier than in the case of $\alpha < 2$, and this indicates the role of LRI 
comparing to the standard case of the nearest-neighbor interaction; (iii)
another important difference of the onset of chaos in $\alpha$DNLS and 
in DNLS can be deduced from the ($Im\ z$, $Re\ z$) plots; in Fig.\ 7 (top right panel) 
for DNLS, trajectory fills space more-or-less uniformly what is typical for 
Hamiltonian chaos, while in Figs. 3, 4 for $\alpha$DNLS trajectories looks like 
in the case of stochastic attractors what is more natural for $\alpha < 2$ \cite{Zas06}. 
There exists an inner part of the diagram that is avoided by the trajectories, 
at least for the observed time.  

The last considered case is the asymmetric initial condition given by 
Eq.\ (\ref{con2}). Numerical solution of the equations of motion reveals 
a difference in this case compared to the symmetric initial condition case. 
For some values of the excitation number $M$ the numerical solution starts 
to move in the left or right dirrection. This dirrection can also change 
randomly in time. Figure 8 shows an example of such behavior for 
$\alpha=1.11$, $J/J_0=1$ and $M=22.22$. 
Note, that the power spectrum in this case is also broad. 


\section{Conclusion}

One of the main feature of the considered $\alpha$DNLS model is implementation of a new, 
additional to the standard DNLS, parameter $\alpha$ that in the continuous limit implies 
the fractional dynamics described by the FNLS. From another point of view, $\alpha$ is 
responsible for stong correlations between distant oscillators, i.e. 
a long-range interaction is introduced through the parameter $\alpha$. This feature of the 
$\alpha$DNLS brings a new physics with a new control parameter. The role of the LRI was known 
before for collective phenomena in complex medium such as chemical or biological set of objects \cite{Kuramoto}, 
phase transition in one-dimentional systems \cite{Dyson}, 
synchronization \cite{Tar05}, regularization in  quantum field theory \cite{Gol04}. Our detailed 
analysis helps to understand some specific properties of destabilization and onset of chaos in 
$\alpha$DNLS with $\alpha < 2$. Similar analysis can be performed for other models with LRI. 
An important part of our analysis is utilization of the possibility to transfrom the behavior of 
discrete chain of interaction objects into the continuous medium equation with the fractional derivatives. 
This formal procedure raise the question of the discrete-continuous equivalence up to a new level where the 
appearance of an additional parameter $\alpha$ increases the difficulty the answer the equation. 


\section*{Acknowledgments}

The authors are grateful to V.E. Tarasov for valuable comments and for the reading of the manuscript. 
This work was supported by the Office of Naval Research,
Grant No. N00014-02-1-0056 and the NSF
Grant No. DMS-0417800.

\end{document}